\documentclass[reprint,
superscriptaddress,
amsmath,amssymb,aps,showkeys,showpacs,
twoside,final,secnumarabic,
nofootinbib]{revtex4-2}

\usepackage[paperwidth=205mm,paperheight=290mm,top=17mm,bottom=25mm,
inner=17mm,outer=17mm,
twoside]{geometry}

\usepackage{cmap} 
\usepackage[T1,T2A]{fontenc}
\usepackage[utf8]{inputenc}
\usepackage[russian,english]{babel}
\usepackage{color}
\usepackage{graphicx}
\usepackage{dcolumn}
\usepackage{bm} 
\usepackage[unicode=true,colorlinks=true,linkcolor=magenta, urlcolor=blue, citecolor = blue,breaklinks]{hyperref}
\usepackage{multirow}
\usepackage{url}
\usepackage{breakurl}
\DeclareGraphicsExtensions{.eps}

\newcount\issue
\newcount\Vol
\newcount\numb
\usepackage{fancyhdr} 
\newcommand{\beq}{\begin{equation}}
\newcommand{\eeq}{\end{equation}}
\newcommand{\bea}{\begin{eqnarray}}
\newcommand{\eea}{\end{eqnarray}}

\newcommand{\eq}{\begin{equation}}
\newcommand{\en}{\end{equation}}
\newcommand{\tr}{\mbox{Tr}}

\def\Vol{\textbf{78}}
\def\numb{x}
\begin{document}

\title{
New results on gauge field decomposition in  $SU(3)$ gluodynamics}

\def\addressa{Institute for High Energy Physics NRC Kurchatov Institute, 142281 Protvino, Russia}
\def\addressb{Pacific Quantum Center, Far Eastern Federal University, 690922 Vladivostok, Russia}

\author{\firstname{V.G.}~\surname{Bornyakov}}
\email[E-mail: ]{Vitaly.Bornyakov@ihep.ru}
\affiliation{\addressa}
\affiliation{\addressb}
\author{\firstname{V.A.}~\surname{Goy}}
\affiliation{\addressb}
\author{\firstname{I.E.}~\surname{Kudrov}}
\affiliation{\addressa}

\received{xx.xx.2026}
\revised{xx.xx.2026}
\accepted{xx.xx.2026}

\begin{abstract}
We study decomposition of the nonabelian gauge field
into the Abelian component created by Abelian monopoles and the modified nonabelian components with monopoles
removed after fixing the Maximal Abelian gauge in $SU(3)$ lattice gluodynamics.
We compute the static potential $V(r)$ for the original gauge field and for its components $V_{mon}$ and $V_{mod}$ at two values of the 
lattice spacing. We confirm that with optimal gauge fixing the sum $V_{mon} + V_{mod}$ deviates substantially from
$V(r)$. We show that this decomposition of the static potential is satisfied with good precision at all distances
when we use another set of Gribov copies.  
\end{abstract}

\pacs{11.15.Ha; 12.38.Gc; 12.38.A}\par
\keywords{lattice QCD, confinement, maximal Abelian gauge, Gribov copies   \\[5pt]}

\maketitle
\thispagestyle{fancy} 


\section{Introduction}\label{intro}
We study properties of the maximal Abelian gauge (MAG) in $SU(3)$ gluodynamics using lattice regularization. MAG was introduced in \cite{thooft2} within the scope of the dual superconductor scenario of confinement proposed by G.~t'Hooft \cite{thooft} and S.~Mandelstam 
\cite{mandelstam}, see Refs.~\cite{Greensite:2003bk,Suganuma:2023mml,Chernodub:2022oza} for reviews. For $SU(3)$ gauge group the gauge is defined by minimization of the functional
\beq
F_{MAG} =  \frac{1}{12V} \int d^4 x  \sum_{\mu, a \neq 3,8} A^a_\mu(x) A^a_\mu(x),
\label{functional}
\eeq 
invariant under Abelian gauge transformations $g(x) \in U(1) \times U(1)$. 
For given gauge field the extrema of this functional (we consider only its minima) are Gribov copies \cite{Gribov:1977wm}: gauge field 
configurations connected by a gauge transformation and satisfying the gauge condition in the differential form:
\begin{equation}
\sum\limits_{b \neq 3, 8}(\partial_{\mu} \delta^{ab} - g f^{ab3} A_\mu^3(x) - g f^{ab8} A_\mu^8(x)) A_\mu^b(x) =0
\end{equation}
for $a \neq 3, 8$.
The respective gauge functional in lattice regularization is 
\begin{equation}
F^{lat}_{MAG} = 1-\frac{1}{12\,V}
 \sum_{x,\mu,a=3,8} \tr ~[U_\mu(x) \lambda_a U^\dagger_\mu(x) \lambda_a ] \,,
\label{functional2}
\end{equation} where $U_\mu(x) \in SU(3)$ is the lattice gauge field defined on lattice links.

It was demonstrated by numerical results that in this gauge the monopole component $A^{mon}_\mu(x)$ of the gauge field (to be defined below) is responsible for the string tension \cite{suzuki,stack} supporting the idea of the dual superconductor scenario. 

In \cite{Bornyakov:2005hf} it was found in $SU(2)$ gluodynamics that decomposition of the gauge field into the monopole component $A^{mon}_\mu(x)$  and the component with monopoles removed (modified component) $A^{mod}_\mu(x)$ 
\beq
A_\mu(x)= A^{mon}_\mu(x) + A^{mod}_\mu(x)
\label{field_decomp}
\eeq
gives rise to the approximate decomposition of the static potential
\beq
V(r) \approx V_{mon}(r) + V_{mod}(r) \,.
\label{pot_decomp}
\eeq
and the monopole term $V_{mon}(r)$ is reproducing the linear
part of the static potential $V(r)$, while the modified term $V_{mod}(r)$ is purely
Coulombic.
We demonstrated in \cite{Bornyakov:2021enf} that decomposition (\ref{pot_decomp}) improves when the lattice spacing decreases and might become exact in the continuum limit. 

The case of $SU(3)$ gluodynamics was considered first in \cite{BornyakovVG:2023rci}. It was shown for one value of the lattice spacing that decomposition (\ref{pot_decomp}) works well at short distances while at large distances the rhs of (\ref{pot_decomp}) deviates substantially from its lhs due to small string tension in $V_{mon}(r)$. In this note, we present results for two lattice spacings and show that proper choice of Gribov copies allows to improve substantially the  precision  of relation (\ref{pot_decomp})  .

\section{Definition of the gauge field decomposition}

We first make an Abelian projection of the gauge field configurations fixed to MAG. This means decomposing the non-Abelian gauge field $U_\mu(x) \in SU(3)$ into the product of the non-diagonal component $U^{offd}_\mu(x) \in SU(3)/U(1) \times U(1)$ and the diagonal component $U^{Abel}_\mu(x) \in U(1) \times U(1)$  \cite{Brandstater:1991sn}:
\begin{equation}
    U_\mu(x) = U^{offd}_\mu(x) U^{Abel}_\mu(x) \,.
    \label{eq:coset}
\end{equation}
The Abelian field $U^{Abel}_\mu(x)$ is determined as
\beq
\label{uabel}
U^{Abel}_\mu(x) = \mbox{diag} \left(u^{1}_\mu(x),u^{2}_\mu(x),u^{3}_\mu(x)
\right)\,,
\eeq
where
\beq
\label{ulink}
u^{a}_{\mu}(x)=e^{i \theta^{a}_{\mu}(x)}
\eeq
with
\beq
\label{tlink}
\theta^{a}_{\mu}(x) = \arg~(U_\mu^{aa}(x))-\frac{1}{3} \sum_{b=1}^3
\arg(U_\mu^{bb}(x))\,\big|_{\,{\rm mod}\ 2\pi}
\eeq
such that
\beq
\theta^{a}_{\mu}(x) \in [-\frac{4}{3}\pi, \frac{4}{3}\pi]\,.
\eeq
This definition of Abelian projection $U_\mu^{Abel}(x)$ maximizes the expression
$|\mbox{Tr} \left( U_\mu^\dagger(x) U_\mu^{Abel}(x) \right) |^2$.
The Abelian gauge field components $\theta_{\mu}^{a}(x)$ can in turn be decomposed into
monopole (singular) and photon (regular) parts:
\beq
\theta_{\mu}^{a}(x)=\theta^{a,\,mon}_{\mu}(x)+\theta^{a,\,ph}_{\mu}(x)\,,
\label{monph}
\eeq
The monopole part is defined by \cite{Smit:1989vg}:
\beq
\theta^{a,\,mon}_{\mu}(x)= 2 \pi \sum_{y} {D(x-y)\partial^-_{\alpha}
m_{\alpha\mu}^{a}(y)}\,, \label{montheta}
\eeq
where integers  $m_{\mu\nu}^{a}(x)$ denote the singular part of the
Abelian plaquettes (Dirac plaquettes),  $\partial^-_{\alpha}$ is the backward lattice derivative, and $D(x)$ denotes the lattice Coulomb propagator. The lattice version of the gauge field decomposition (\ref{field_decomp}) is
\beq
U_\mu(x) = U_\mu^{mod}(x)\, U_\mu^{mon}(x)\,,
\eeq
where $U^{mon}_\mu(x)$ is defined as
\beq
\label{umon}
U^{mon}_\mu(x) = \mbox{diag} \left(e^{i\theta^{1,\,mon}_\mu(x)},e^{i\theta^{2,\,mon}_\mu(x)},e^{i\theta^{3,\,mon}_\mu(x)} \right)\,.
\eeq

\section{Numerical results for the static potential decomposition}

We fixed MAG using the simulated annealing (SA) algorithm which is the optimal choice for MAG, as well as for other gauges that require minimization of the gauge fixing functional. This algorithm was first used to fix MAG in the $SU(2)$  case \cite{Bali:1994jg} and then
was extended to the $SU(3)$  group in \cite{Bornyakov:2001qw}. A less powerful algorithm is the relaxation plus overrelaxation (RO) algorithm. 

Below we present results obtained with the use of these two algorithms. 
For SA algorithm we produced 20 Gribov copies per configuration and chose the one with the lowest value of the functional (\ref{functional2}). For RO algorithm only one gauge copy was produced. The values of the functional 
(\ref{functional2}) are substantially lower in the first case. Similarly the monopole density is lower.  The respective numerical values were presented in our earlier
publications  \cite{Kudrov:2024efi,Kudrov:2025nso} devoted to the influence of the Gribov copies on the Abelian and monopole dominance.
We used the implementation of the gauge fixing algorithms presented in \cite{Schrock:2012fj}. 
The corresponding code can be downloaded from \text{https://github.com/havogt/culgt/tree/master}.

The numerical simulations were completed using the Wilson lattice action at two values of the lattice coupling constant $\beta=6/g^2$:
$\beta=6.0$ and 6.1, corresponding to lattice spacings $a=0.93$~fm and 0.79~fm. We computed $r/a \times t/a$ rectangular Wilson loops $W(r/a,t/a)$, $W_{mon}(r/a,t/a)$ and $W_{mod}(r/a,t/a)$ using lattice gauge fields $U_\mu(x)$,  $U_\mu^{mon}(x)$, $ U_\mu^{mod}(x)$ introduced above. The APE smearing \cite{APE:1987ehd} and hypercubic blocking \cite{Hasenfratz:2001hp}
algorithms were employed to extract the respective static potentials $V(r)$, $V_{mon}(r)$, and $V_{mod}(r)$.

\begin{figure*}
\includegraphics[width=95mm]{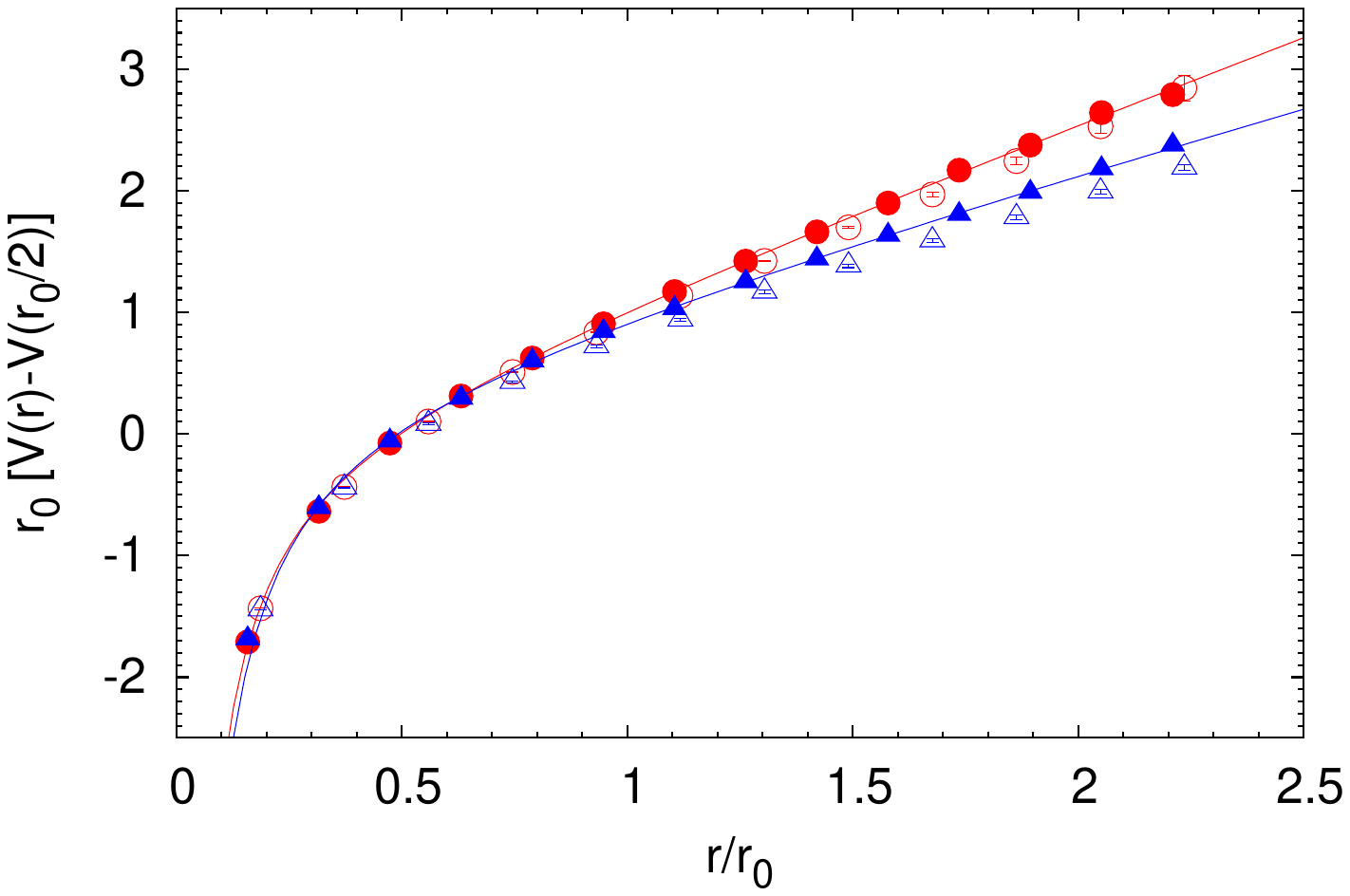}%
\hspace*{-1cm}
\includegraphics[width=95mm]{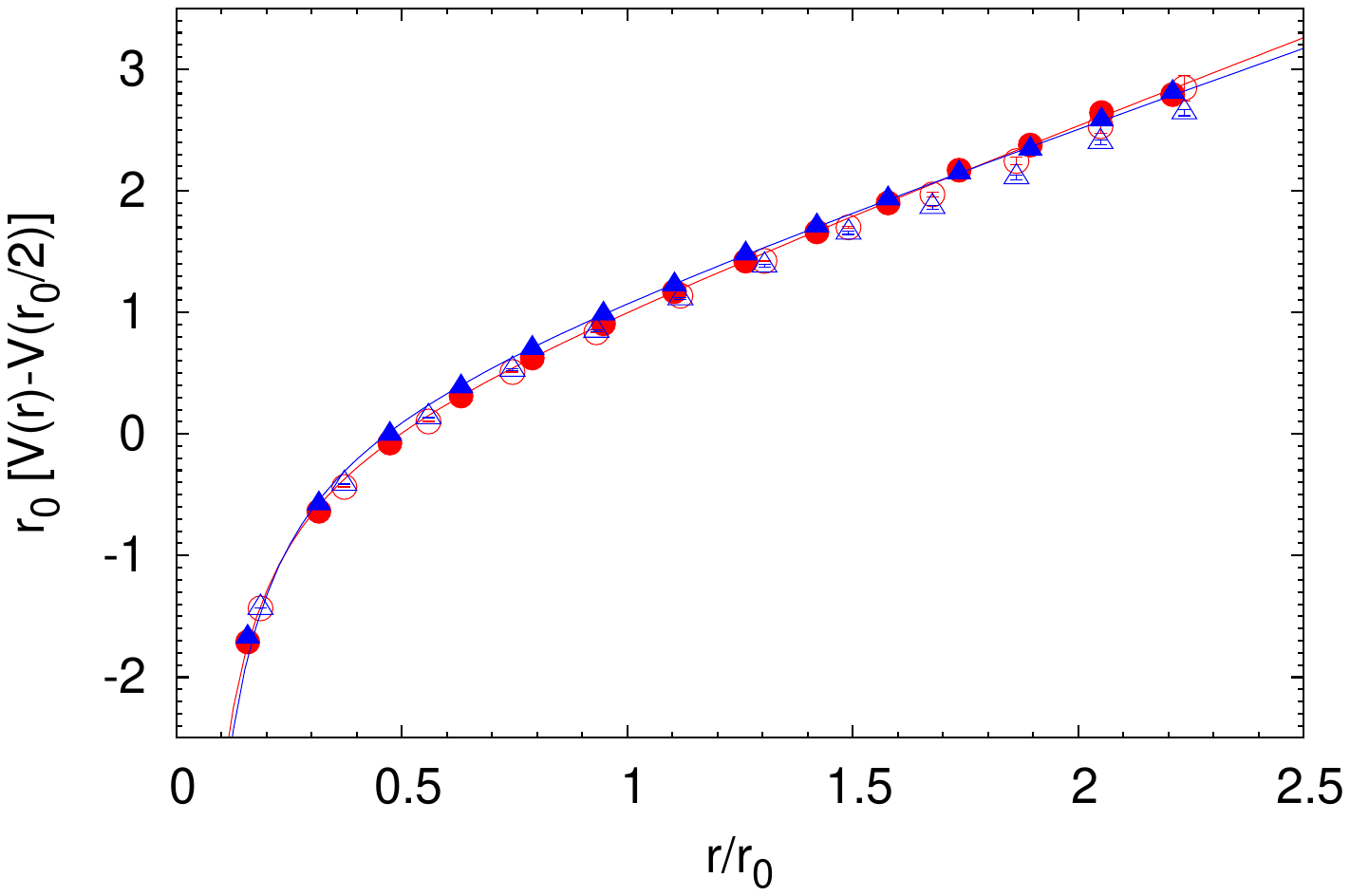}%
\vspace{-1cm}
\caption{\label{fig:potential} Comparison of the static potential $V(r)$ computed for original gauge field (circles) with  $V_{mon}(r)+V_{mod}(r)$ (triangles) for two couplings: $\beta=6.0$ (empty symbols) and $\beta=6.1$ (filled symbols). Left: $V_{mon}(r)+V_{mod}(r)$ was obtained with SA algorithm; Right: $V_{mon}(r)+V_{mod}(r)$ was obtained with OR algorithm. }
\label{fig:potentials1}
\end{figure*}

In Fig.~\ref{fig:potentials1} our results are presented. The potentials are shifted by $V(r_0/2)$ where $r_0=0.5$~fm is a Sommer parameter. In the left figure results for $V_{mon}(r)+V_{mod}(r)$ obtained with SA algorithm are presented. We confirm our earlier result \cite{BornyakovVG:2023rci} that $V_{mon}(r)+V_{mod}(r)$ deviates substantially at large distances due to small string tension in $V_{mon}(r)$. Comparing results for two lattice spacings we conclude that this behavior is independent of the lattice spacing.
In the right figure we present results for another set of Gribov copies obtained with RO algorithm. It is clear that these Gribov copies produce monopole and modified components which satisfy decomposition (\ref{pot_decomp}) with good precision at all distances. This is a main result of our work. It is still necessary to study further effects of lattice spacing and Gribov copies on precision of decomposition (\ref{pot_decomp}). This work is in progress.

\begin{figure}
\vspace{-1cm}
\hspace*{-.5cm}
\includegraphics[width=95mm]{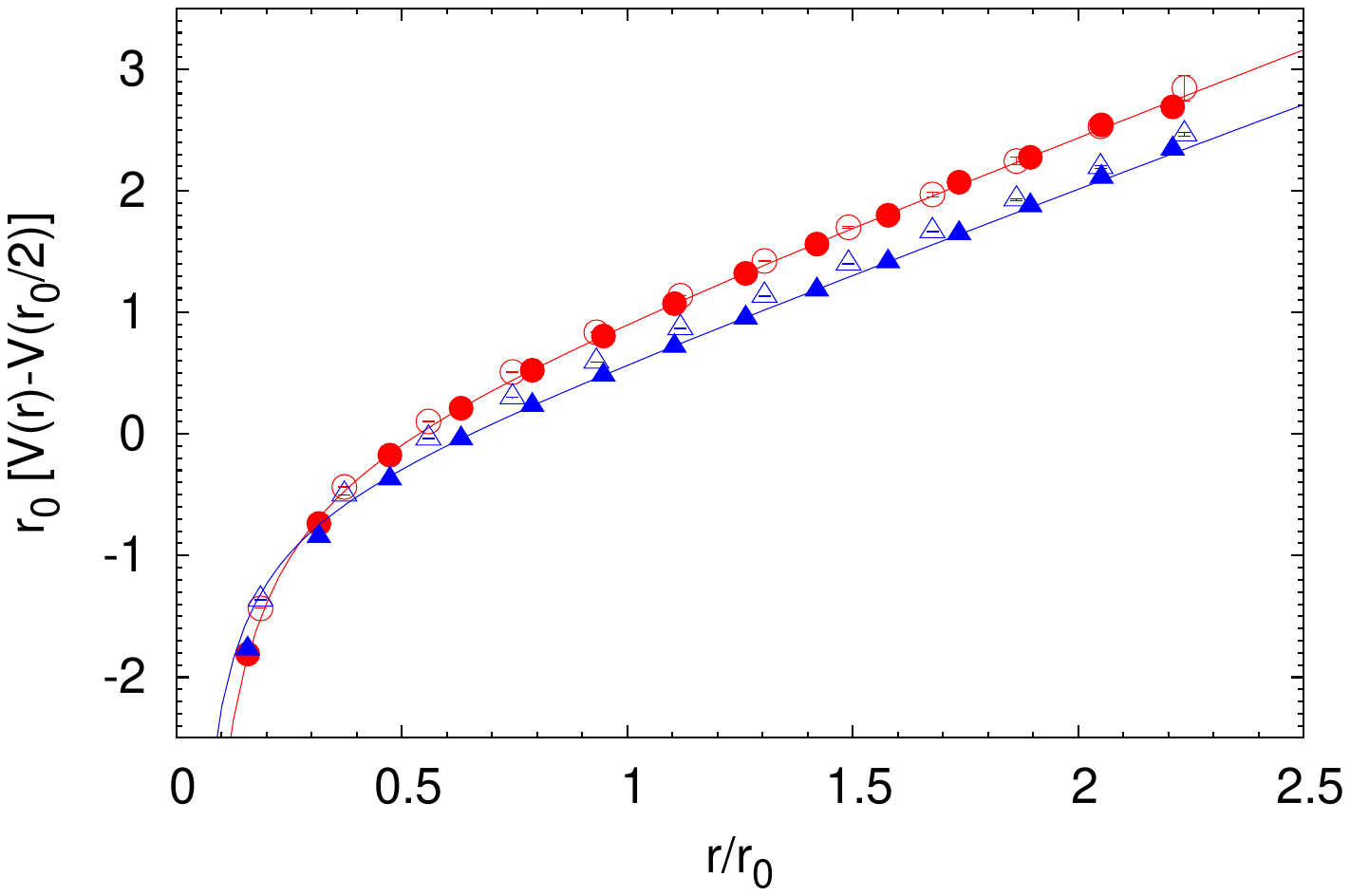}%
\vspace{-1cm}
\caption{\label{fig:potential} Comparison of the static potential $V(r)$ (circles) with  $V_{offd}(r)+V_{Abel}(r)$ (triangles) for two couplings: $\beta=6.0$ (empty symbols) and $\beta=6.1$ (filled symbols). Results for $V_{offd}(r)$ and $V_{Abel}(r)$  were obtained with OR algorithm. }
\label{fig:potentials2}
\end{figure}

The decomposition different from eq.~(\ref{pot_decomp}) 
was suggested in  \cite{Sakumichi:2014xpa}.
 The usual coset decomposition (\ref{eq:coset})  was used
and respective decomposition for the static potential was verified:
\begin{equation}
 V(r) \approx  V_{offd}(r) + V_{Abel}(r).
\label{eq:poten_decomp2}
\end{equation}
We show our results for decomposition (\ref{eq:poten_decomp2}), computed on Gribov copies obtained with RO algorithm, in Fig.~\ref{fig:potentials2}. It can be seen that there is no agreement between lhs and rhs of (\ref{eq:poten_decomp2}). It is clear that the reason for discrepancy is the Coulomb term while the string tension in $V_{offd}(r) + V_{Abel}(r)$ is very close to the physical string tension in agreement with results of Ref.~\cite{Sakumichi:2014xpa}.

\section{Conclusions}
We continued our study of the gauge field decomposition  into the monopole
and the modified (monopoleless) components  after fixing MAG. Here
we presented results for two lattice spacings and for two sets of Gribov copies for each
lattice spacing. With effective SA algorithm we obtained Gribov copies which do not support the decomposition (\ref{pot_decomp}) at large distances due to small string tension in $V_{mon}$ term, independent of the lattice spacing. Then, we demonstrated that on Gribov copies obtained with RO algorithm the  decomposition (\ref{pot_decomp}) is satisfied quite precisely at all distances and for both lattice spacings.

\section*{FUNDING}
VGB and VAG have been supported by Grant No. FZNS-2024-0002 of the Ministry of Science and Higher Education of Russia.

\section*{CONFLICT OF INTEREST}
The authors declare that they have no conflicts of interest.

\nocite{*}


\end{document}